\documentclass[
superscriptaddress,
amsmath,
amssymb,
reprint,%
author-year,
]{revtex4-1}
\usepackage{amsmath}
\usepackage{mathtools}
\usepackage{graphicx}
\usepackage{dcolumn}
\usepackage{bm}
\usepackage{algpseudocode}
\graphicspath{{figures/}}
\usepackage{subfigure}

\usepackage{algorithm}
\usepackage{algpseudocode}
\usepackage[utf8x]{inputenc} 
\usepackage{xspace}
\usepackage{latexsym}
\usepackage{array}
\usepackage{fancybox,framed}
\usepackage{graphicx}

\usepackage{natbib}
\usepackage{bm}
\usepackage{xcolor}
\usepackage{qcircuit}
\usepackage{braket}
\usepackage{float}
\usepackage{algpseudocode}
\usepackage{hyperref}%

\usepackage{mathtools}
\usepackage{siunitx}
\begin{document}


\title{Unentangled quantum reinforcement learning agents in the OpenAI Gym}%

\author{Jen-Yueh Hsiao}
\email{auston.jy.hsiao@foxconn.com}
\affiliation{Department of Physics and Center for Theoretical Physics, National Taiwan University, Taipei 10617, Taiwan}
\affiliation{Hon Hai (Foxconn) Research Institute, Taipei, Taiwan}


\author{Yuxuan Du}
\affiliation{JD Explore Academy, Beijing 101111, China}
\author{Wei-Yin Chiang}%
\affiliation{Hon Hai (Foxconn) Research Institute, Taipei, Taiwan}
\author{Min-Hsiu Hsieh} \email{min-hsiu.hsieh@foxconn.com}
\affiliation{Hon Hai (Foxconn) Research Institute, Taipei, Taiwan}

\author{Hsi-Sheng Goan} \email{goan@phys.ntu.edu.tw}
\affiliation{Department of Physics and Center for Theoretical Physics, National Taiwan University, Taipei 10617, Taiwan}
\affiliation{Center for Quantum Science and Engineering, National Taiwan University, Taipei 10617, Taiwan}
\affiliation{Physics Division, National Center for Theoretical Sciences, Taipei, 10617, Taiwan}

\date{\today}

\begin{abstract}
Classical reinforcement learning (RL) has generated excellent results in different regions {; however, its sample inefficiency remains a critical issue.} {In this paper, we provide concrete numerical evidence that the sample efficiency {(the speed of convergence)} of quantum RL could be better than that of classical RL, {and for achieving comparable learning performance, quantum RL could 
use much (at least one order of magnitude) fewer trainable parameters than classical RL}. Specifically, we employ the popular benchmarking {environments of RL in the} OpenAI Gym, and show that our quantum RL agent converges faster than classical fully-connected neural networks (FCNs) in the tasks of CartPole and Acrobot under the same optimization process. We also successfully train the first quantum RL agent that can complete  the task of LunarLander in the OpenAI Gym. Our quantum RL agent only requires {a single-qubit-based variational quantum circuit} without entangling gates, followed by a classical neural network (NN) to post-process the measurement output. Finally, we could accomplish the aforementioned tasks on the real IBM quantum machines. To the best of our knowledge, none of the earlier quantum RL agents could do that.}
\end{abstract}


\maketitle
\section{\label{sec:1}Introduction}

Classical reinforcement learning (RL)~\cite{10.5555/3312046} has generated excellent results in different regions~\cite{silver2017mastering, silver2017mastering, Vinyals2019, Wurman2022, Degrave2022, panou2020deepfoldit, mirhoseini2020chip}. During the past decade, RL has been broadly applied to master Go~\cite{silver2017mastering}, design chips~\cite{mirhoseini2020chip}, play the game for StarCraft and Gran Turismo~\cite{Vinyals2019, Wurman2022}, improve the nuclear fusion problem~\cite{Degrave2022}, and solve the problem of protein folding~\cite{panou2020deepfoldit}. Despite the remarkable achievements, most RL techniques fail to balance the tradeoff between exploitation and exploration~\cite{rlblogpost}. The difficulty comes from the fact that the state-action space is exponentially large~\cite{du2020agnostic}, where the optimal policy can not be explored efficiently. A mainstream strategy is to feed numerous trials to RL models in the 
optimization process to enhance performance~\cite{nachum2018dataefficient, ye2021mastering, whitney2021decoupled, badia2020up, liu2020data, zhang2020sample, agarwal2020theory, bhandari2020global}. Nevertheless, such a sample inefficiency challenges the applicability of RL towards large-scale problems~\cite{rlblogpost}, where the requested computational overhead is expensive or even unaffordable. For example, in the task of playing an Atari game, two representative RL models, i.e., Deep Q-learning~\cite{Mnih2015} and Rainbow RL~\cite{hessel2017rainbow}, achieve good performance after about 80 and 300 hours of play experience in an Atari game, while humans learn it within a few minutes. In this regard, improving sample efficiency (the speed of convergence) is the key to using RL to solve complex real-world problems.

Quantum computing targets to achieve certain computational advantages beyond the reach of classical computers~\cite{Arute2019, 2020_quantum_advantage, 2021_strong_quantum_computational, PhysRevLett.119.170501}. In the noisy intermediate-scale quantum (NISQ) era~\cite{Preskill2018quantumcomputingin, bharti2021noisy}, a promising candidate for this goal is quantum machine learning (QML)~\cite{Biamonte2017}. QML models can mainly be  categorized into quantum supervised learning~\cite{18, 19, PhysRevResearch.3.033056, chen2020hybrid_tensornetwork, 2021_endtoend, Henderson2020, chen2021hybrid, chen2020quantum_convolution, Wang2020, Du_2021, chen2020quantum, 9413453}, quantum unsupervised learning \cite{Zoufal2019,PhysRevApplied.16.024051, rudolph2021generation, 2019_Training_of_quantum_circuit, du2021exploring, 2018_variational_autoencoder,PhysRevResearch.2.033125}, and quantum reinforcement learning (QRL)~\cite{variational_QRL_Chen, Lockwood_Si_2020, Parametrized_quantum_policies_Jerbi, skolik2021quantum, lan2021variational_soft_actor_critic, Evolutionary_Optimization_VQC_chen, kwak2021introduction}. Extensive studies have been conducted to explore the potential advantages of quantum supervised and unsupervised learning models. Concretely, for the synthetic dataset, Refs.~\cite{Huang2021, Liu2021, Wang2021towards} exhibited the advantages of quantum neural networks~\cite{Havlicek2019, articleBeer} and quantum kernels~\cite{Havlicek2019} in the measure of generalization error~\cite{PRXQuantum.2.040337, Abbas2021, PRXQuantum.2.040321, bu2021statistical, du2021efficient, PhysRevLett.126.190505}. However, quantum supervised and unsupervised learning models may encounter trainability issues, where the gradients exponentially vanish for the number of qubits~\cite{2021arXiv211215002Z, PhysRevLett.127.120502}. 
{Moreover, a recent study has shown that, in fact, the performance of quantum supervised learning models on real-world datasets could be worse than that of classical learning models~\cite{qian2021dilemma}.}


\begin{table*}[http]
\centering
  \caption{Related works of VQC-based reinforcement learning in OpenAI Gym.}
  \begin{ruledtabular}
    \begin{tabular}{c c  c  c  p{6em}  c }
      Literature & Environments & Learning algorithm & Solving tasks & Comparing with classical NNs & Using real devices\\
      \hline
       \cite{variational_QRL_Chen}& FrozeLake & Q-learning  & Yes & None  & Yes   \\
        
       \cite{Lockwood_Si_2020}& CartPole-v0, blackjack & Q-learning & No  & Similiar performance  & No \\
       \cite{Parametrized_quantum_policies_Jerbi}& CartPole-v1, Acrobot  &Policy gradient with baseline & No  & None  & No     \\
        \cite{Parametrized_quantum_policies_Jerbi}&  MountainCar  &Policy gradient with baseline & Yes  & None  & No     \\
    \cite{skolik2021quantum}& CartPole-v0, FrozeLake  & Q-learning  & Yes  & None   & No    \\
     \cite{lan2021variational_soft_actor_critic}& Pendulum  & Soft Actor-Critic & Yes & Similar performance & No   \\
    
     \cite{kwak2021introduction}& CartPole-v0  & Proximal policy optimization & No & None & No  \\
     Our work & CartPole-v1, Acrobot, LunarLander  & Proximal policy optimization & Yes & Fast convergence & Yes\\
      \end{tabular}%
  \end{ruledtabular}
  \label{tab_related_workd}
\end{table*}



{Besides the attempts to understand the potential advantages of quantum supervised learning models, there is a growing interest in designing powerful QRL models to compensate for the caveats of classical RL models, such as sample inefficiency. There are not proven theoretic results regarding the advantages of QRL models up to date. Instead, most studies numerically evaluate the performance of their proposals~\cite{Saggio2021, 2008_quantum, 2014_Quantum_Speedup, huang2021quantum, Hamann2021}}. {Concretely, Refs.~\cite{variational_QRL_Chen, Lockwood_Si_2020, Parametrized_quantum_policies_Jerbi, skolik2021quantum, lan2021variational_soft_actor_critic, Evolutionary_Optimization_VQC_chen, kwak2021introduction} have attempted to {improve sample inefficiency} by using multi-qubit variational quantum circuit (MVQC)~\cite{2021_variationalreview, 2018} that has lots of entangling gates on a generic benchmark OpenAI Gym~\cite{1606.01540}. However, none of them outperforms classical RL models~\cite{Lockwood_Si_2020, Parametrized_quantum_policies_Jerbi, skolik2021quantum, kwak2021introduction}.} Alternatively, Refs.~\cite{Saggio2021, 2008_quantum, 2014_Quantum_Speedup, Hamann2021, pmlr-v139-wang21w} designed QRL algorithms based on Grover's algorithms~\cite{grover1996fast, ahuja1999quantum, 2015_quantumSpeedup} to improve the sample efficiency. Unfortunately, these algorithms are hard to implement on NISQ devices. Considering the ambitious aim of QRL is to provide computational advantages over classical NN on real-world tasks, it is natural to ask: ``\emph{Do the current QRL agents surpass the classical RL agent in OpenAI Gym?}" If the response is positive, it is necessary to figure out ``\emph{how would we design the model?}"

\subsection{Main results}

{We demonstrate a series of training and testing processes to address the previous question. We first propose a {single-qubit-based variational quantum circuit (SVQC) model} that only consists of single-qubit rotational gates. Our SVQC models show the better convergence compared to the classical fully-connected neural networks (FCNs) on the learning curves in the tasks of CartPole and Acrobot, and can use much (at least one order of magnitude) fewer trainable parameters than the classical FCNs to accomplish comparable or better learning performance. Furthermore, our SVQC models achieve higher rewards than other VQC-based models in the CartPole and Acrobot tasks~\cite{Lockwood_Si_2020, Parametrized_quantum_policies_Jerbi, skolik2021quantum, Evolutionary_Optimization_VQC_chen}.
While we first successfully train the quantum agent to accomplish the LunarLander task in the QRL field, our trained models also exhibit satisfactory performances in the testing tasks of CartPole-v0, Acrobat-v1 and LunarLander-v2 on the IBM quantum devices. 
}

\subsection{Related work}

This section collects related works, where VQC-based quantum RL agents were used to solve tasks in the OpenAI Gym. For ease of comparison, these results are summarized in Table~\ref{tab_related_workd}.
Specifically, the Frozen Lake environment in toy text tasks was first solved by Chen et al.~\cite{variational_QRL_Chen}. {The CartPole-v0 task was first attempted in Ref.~\cite{Lockwood_Si_2020} with quantum Q-learning, but its performance is not satisfactory.} 
The control tasks of CartPole-v0, MountainCar, and Pendulum were subsequently accomplished in Ref.~\cite{skolik2021quantum, Parametrized_quantum_policies_Jerbi, lan2021variational_soft_actor_critic}.  
The employed learning algorithms in \cite{variational_QRL_Chen, skolik2021quantum, Parametrized_quantum_policies_Jerbi, lan2021variational_soft_actor_critic} were also included in Table~\ref{tab_related_workd}.
{However, whether the VQC-based model can accomplish the more challenging tasks in OpenAI Gym, e.g., CartPole-v1, LunarLander-v2 and box2d, remains to be answered.} 





Finally, we remark that the aforementioned tasks were conducted using ideal simulators. It is unknown whether noisy quantum RL agents could achieve satisfactory performance.



The paper is organized as follows. Preliminary of classical RL and VQC-based QRL are described in Section~\ref{sec:2}. A novel variational QRL with single-qubit is described in Section~\ref{sec:3}. The simulation results and associated discussions are presented in Section~\ref{sec:4s}. 
The concluding remarks and open questions are presented in Section~\ref{sec:5}.







\begin{figure*}[htbp]
  \centering
  \includegraphics[width=1\linewidth]{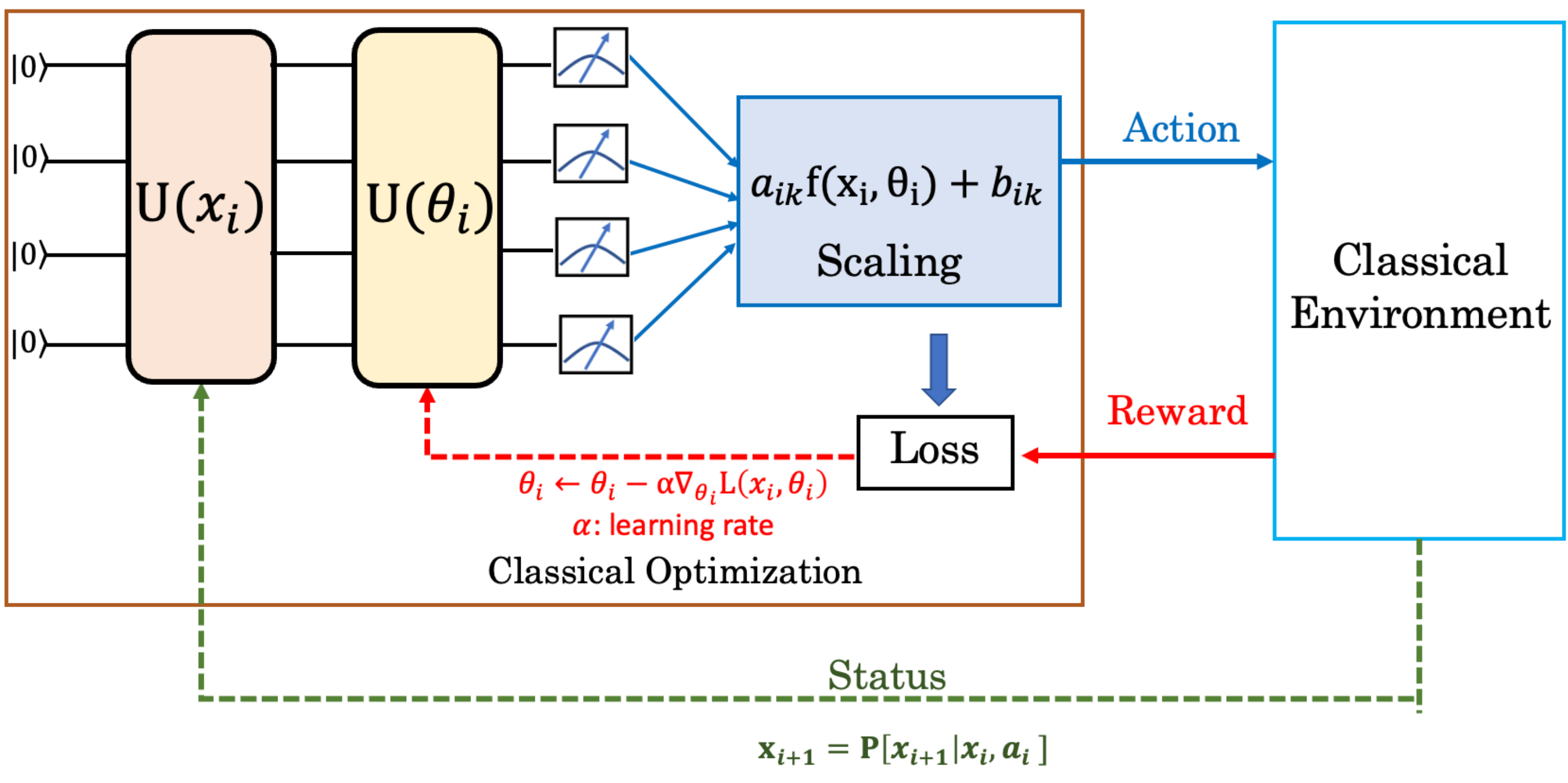}
  \caption{The flow of variational-based quantum reinforcement learning.
  The environment provides the {input} state's features $x_i$ {at step $i$}. The features are encoded to the quantum state by parameterized circuits. Then, the quantum state evolves with the parameters layer $\mathrm{U}(\theta_i)$ and the state is measured by projective {operator} repeatedly. The classical computer calculates the loss function by the {scaled} circuit output based on the measurements. The agent chooses an action $a_{i}$ through the policy, which is dependent on the circuit output, and the agent executes the action on the environment. It receives the reward and next state's features $x_{i+1}$ from the environment. The classical computer calculates the new loss function dependent on the rewards and policy. The trainable parameters $\theta_{i}, a_{i k}, \text{and } b_{i k}$ are {updated} to minimize the loss function.}
  \label{The_flow_of_VQC_base_QRL}
\end{figure*}
\section{\label{sec:2}Preliminary}
Here we briefly recap classical reinforcement learning (RL) in Section~\ref{classical_RL}, RL with variational quantum circuit (VQC) in Section~\ref{variational_quantum_circuit} and introduction to the OpenAI Gym in Section~\ref{sec_introduction_to_openAIgym}.

\subsection{Classical reinforcement learning}\label{classical_RL}

Markov decision process (MDP)~\cite{10.5555/551283} provides a dynamical framework that captures two key features in classical reinforcement learning (RL); namely, \emph{trial and error} as well as \emph{delayed rewards}~\cite{10.5555/3312046}. An MDP can be described as {a 5-tuple,} $\mathcal{M}=(\mathcal{S}, \mathcal{A}, \mathcal{P}, \mathcal{R}, \gamma)$, where $\mathcal{S} \in \mathbb{R}^{d}$ is the state space, $d$ is the dimension of states, $\mathcal{A} \in \mathbb{R}$ is the action space, $\mathrm{P}(x_{i+1}|x_i,a_i) \in \mathcal{P}$ is the probability of transitioning into state $x_{i+1}$ upon taking action $a_i$ in state $x_i$, $(x_i , a_i) \in \left(\mathbb{R}^{d}, \mathbb{R}\right)$ is the state-action pairs at step $i$, $\mathcal{R} \in {[0, R_{\text{max}}]}$ is the  reward space, $R_{\text{max}} \in \mathbb{N}$ is a constant, and $\gamma \in \left[0,1\right)$ is the discount factor~\cite{Bel}. An agent begins at an initial state $x_0$ sampled from an initial distribution $\mathrm{P}(x_0)$. Then it implements the policy $\mathrm{\pi}(a_i|x_i) \in [0,1]$ to take action $a_i \in \mathcal{A}$ at step $i$ from a state $x_i \in \mathcal{S}$ and moves to a next state $x_{i+1} \sim \mathrm{P}(\cdot|x_i,a_i)$. The next state $x_{i+1}$ is dependent on the current state $x_i$ and the agent's action $a_i$. After each action, the agent receives a reward $\mathrm{r}_i = \mathrm{R}(x_i, a_i) \in \mathcal{R}$. Therefore, the relation between action and reward is similar to~\emph{trial and error}, and the reward function $\mathrm{r}_i$ is associated with~\emph{delayed rewards}.

The goal of modern RL with classical neural network (NN) is to maximize the discount expected rewards
\begin{equation}
\mathrm{R}_{i}(\tau)=\hat{\mathbb{E}}_{\pi}\left[\sum_{i=0}^{T} \gamma^{i}\operatorname{r}_{i}(x_i, a_i)\right],  
\end{equation}
where $\tau=(x_0, a_0, x_1, a_1,\dots, x_i, a_i)$ is the trajectory in an episode, $\hat{\mathbb{E}}_{\pi}\left[\dots\right]$ denotes the expectation value under all possible policies. 
For the agent with the state $x_i$ at time $i$, the probability of the agent to take the action $a_i$ is $\mathrm{\pi}_{\theta_{i}}(a_i|x_i)$. The agent learns the stochastic policy $\mathrm{\pi}_{\theta_{i}}(a_i|x_i)$ that is dependent on $x_i$ and $\theta$, where $\theta_{i} \in \mathbb{R}^{m}$ is trainable parameters in classical NN, $m \in \mathbb{N}$ is the dimension of parameters to reach high expected reward. The policy is improved through updating the parameters by gradient ascent $\theta_{i} \leftarrow \theta_{i}+\eta \nabla \operatorname{L}(x_{i},{\theta_{i})}$, where $\operatorname{L}(x_{i},{\theta_{i}) \in \mathbb{R}}$ is the loss function, $\eta \in \mathbb{R}$ is learning rate.

Defining the loss function plays a {crucial} part in optimization problems. 
The loss function of the PPO-clip is as follows:

\begin{equation}
\begin{aligned}
    &\operatorname{L}(x_i, \theta_{i})\\
    = &\hat{\mathbb{E}}_{\pi}\left[\min \left(\operatorname{r}_{i}(\theta_{i}) \operatorname{\hat{A}}_{i}, \operatorname{clip}\left(\operatorname{r}_{i}(\theta_{i}), 1-\epsilon, 1+\epsilon\right)
    \operatorname{\hat{A}}_{i}\right)\right]\label{PPOLoss_function},
\end{aligned}
\end{equation}
where $\operatorname{r}_{i}(\theta_{i})=\frac{\operatorname{\pi}_{\theta_{i}}(a_i|x_i)}{\operatorname{\pi}_{\theta^{\prime}_{i}}(a_i|x_i)}$ is the ratio of new and old policy, ${\theta^{\prime}_{i}}$ is the policy parameters before the update, $\epsilon \in \mathbb{R}$ is usually a small number.
PPO-clip is an {robust} algorithm in various experimental tests~\cite{schulman2017proximal}. 
Moreover, it is {commonly used} in RL algorithms for the OpenAI Gym because it is {easily} operated with good performance.


\subsection{Reinforcement learning using variational quantum circuit  }\label{variational_quantum_circuit}

{The VQC-based QRL is illustrated in Fig.~\ref{The_flow_of_VQC_base_QRL}.}
The key concept of VQC-based QRL~\cite{variational_QRL_Chen, Lockwood_Si_2020, Parametrized_quantum_policies_Jerbi, skolik2021quantum, lan2021variational_soft_actor_critic, Evolutionary_Optimization_VQC_chen, kwak2021introduction} is to learn the policy to acquire the maximum expected rewards $\mathrm{R}_{i}(\tau)=\hat{\mathbb{E}}_{\pi}\left[\sum_{t=0}^{T} \gamma^{i}\operatorname{r}_{i}(x_i, a_i)\right]$ by replacing the classical NN with VQC. 


First, the input data features $x_i$ {at step $i$} from the environment are transferred as $|\Psi_{\text{in }}(x_{i})\rangle=\operatorname{U}(x_{i})|0\rangle$, where $|0\rangle \in \mathbb{C}^{2^n}$ is the initial state, $n \in \mathbb{N}$ is the number of qubit, $|\Psi_{\text{in}}(x_{i})\rangle \in \mathbb{C}^{2^n}$ is the quantum state after encoding the input $x_{i}$, and the operator  $\operatorname{U}(x_{i})$ is an unitary dependent on $x_{i}$. Then, the quantum state $|\Psi_{\text{in}}(x_{i})\rangle$ evolves by the operator $\operatorname{U}(\theta_{i})$, where $\operatorname{U}(\theta_{i})$ is an unitary, $\theta_{i} \in \mathbb{R}^{m}$ is the trainable parameters in VQC, $m \in \mathbb{N}$ is the dimension of parameters. {The resultant quantum state} is measured by projective operator. A general form of circuit output is 
\begin{equation}
\mathrm{f}(x_{i}, {\theta}_{i})=\left\langle 0^{\otimes n}\left|\operatorname{U}^{\dagger}(x_{i}, \theta_{i}) \operatorname{M} \operatorname{U}(x_{i}, \theta_{i})\right| 0^{\otimes n}\right\rangle,   
\label{general_circuitoutput}
\end{equation}
where $\operatorname{U}(x_{i},\theta_{i}) \in \mathbb{M}_{2^n}$
is an unitary operator that depends on the input and trainable parameters, and $\operatorname{M} \in \mathbb{M}_{2^n}$ is a projective operator.

Second, the circuit output based on the measurement is scaled linearly by parameters, $\mathrm{y}_{i p}^{\prime}=a_{i p}\times\mathrm{f}(x_{i}, \theta_{i})+b_{i p}$, where $a_{i p}, b_{i p} \in \mathbb{R}$ are the trainable parameters at {step $i$}, where {$p$} is the index of actions.

The agent's policy decides the probability of {action} $a$ depending on {the scaled} output
\begin{equation}
    \mathrm{P}_{\theta_{i}}(a_{i}|x_{i})=
   \pi_{\theta_{i}}(a_{i}|x_{i}) = \operatorname{Softmax}\left(\mathrm{y}_{i p}^{\prime}\right)=\frac{\mathrm{e}^{\mathrm{y}_{i p}^{\prime}}}{\sum_{p=1}^{k} \mathrm{e}^{\mathrm{y}_{i p}^{\prime}}},
\end{equation}
where $a_{i}$ is the action at the state $x_{i}$, {and $k$} is the number of {actions}. After interacting with the environment, the agent receives the reward and next state $x_{i+1}$. The loss functions $\operatorname{L}(x_{i}, \theta_{i}) \in \mathbb{R}$ are dependent on the {scaled} output and the cumulative rewards. Finally, all trainable parameters, namely $\theta_{i}, a_{i p}, \text{and } b_{i p}$ are optimized by gradient descent on a classical optimizer.

\begin{figure*}[http]
  \centering
    \includegraphics[width=0.7\linewidth]{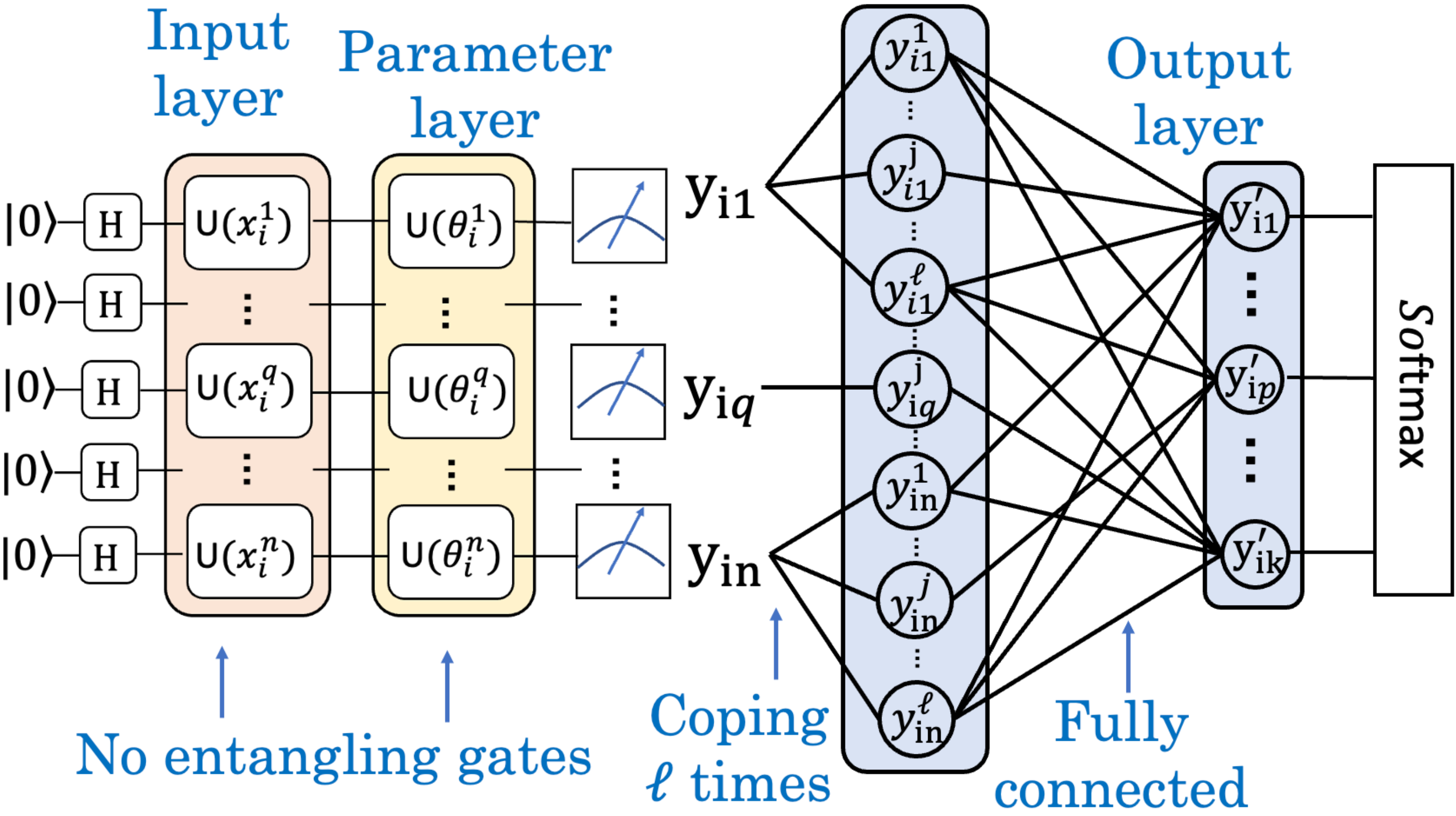}
    \caption{The architecture composes of three parts: input, parameter, and output layers. For input layer, the environment provides $n$-dimensional state {with  $x_i^q$ being the input state of qubit $q$, and for the  CartPole, Acrobot, and LunarLander environments, the dimension of the input state also corresponding to the number of qubits are $n$ = 4, 6, and 8, respectively. The input state $x_i^q$ is encoded by the angle of a single-qubit rotational gate $\mathrm{U}(x_i^q)$. In the parameters layer, we use the trainable parameters $\theta_i^q$ to control single-qubit gates and the circuit has no entangling gates}. In the output layer, {the output of each qubit in the circuit is reused (copied) $\ell$ times and then all outputs are connected with classical NN.}
    {$\mathrm{y}^{j}_{i q}$ is the $j\text{th}$ copy of the output of the $q\text{th}$ qubit at step $i$, $\mathrm{y}_{i q}$. The duplicated outputs $\mathrm{y}^{j}_{i q}$ with $q=1,\dots,n$ and $j=1,\dots,\ell$ are fed into the classical fully-connected NN to produce the scaled circuit outputs $\mathrm{y}_{i p}^{\prime}$ with $p=1,\dots,k$. Finally, we add the Softmax function to transfer the scaled circuit outputs to probability distribution.}} 
    \label{Fig_ReuseCircuit}
\end{figure*}

\begin{table}[htbp]
  \caption{The {performance metrics} {for several OpenAI Gym environments.}}
  \begin{ruledtabular}
    \begin{tabular}{c p{4em} p{5em} p{8em} }
     Environment &  Number of status & Number of actions &The metric of performance\\   
    \hline
     CartPole-v0 & 4 & 2 & Average reward of 195.0 over 100 consecutive trials.\\
     CartPole-v1 & 4 & 2 & Average reward of 475.0 over 100 consecutive trials. \\
     Acrobot-v1  & 6 & 3 & Do not define "solving" condition. Look at the $\href{https://github.com/openai/gym/wiki/Leaderboard}{\text{Gym} \text{ }\text{leadboard}}$ to evaluate the model. \\
     LunarLander-v2 &8 &4& Average reward of 200 over 100  consecutive trials. 
     \end{tabular}
  \end{ruledtabular}
  \label{tab_metric_performance}
\end{table}

\subsection{\label{sec_introduction_to_openAIgym}OpenAI Gym environments}

OpenAI Gym provides benchmarking environments for RL tasks to compare their model performance. 
CartPole, Acrobot, and LunarLander tasks are regarded as the basic environments in OpenAI Gym.
The schematic diagrams and the {performance metrics which measure how well RL agents can achieve the intended goals}
of these three tasks are shown in Fig.~\ref{fig_openAIresult} (a)-(c) and Table~\ref{tab_metric_performance}{, respectively}. 

The goal of the CartPole task is to balance the pole on a cart by moving the cart. A reward of $+1$ is {given} for each step that the pole remains upright.
The episode ends when the pole is more than 12 degrees from vertical, or the cart moves more than 2.4 units from the center.

Acrobot-v1 is another classical control task in the Gym. The system includes two joints and two links, where the joint between the two links is actuated. Initially, the links are hanging downwards, and the goal is to swing the end of the lower link up to a given height.


LunarLander is a more complex task than CartPole and Acrobot tasks. Its goal is to let the agent learn to land between the two yellow flags. The agent controls up to four actions corresponding to no action, the main engine is firing down, and the engine is firing left or right.

{More descriptions about the CartPole, Acrobot and LunarLander environments together with discussions on input data encoding schemes, measurements and VQC-based quantum RL can be found in Appendix~\ref{Appendix:VQC_Environments}.}

\section{\label{sec:3}A novel variational quantum reinforcement learning with single-qubit}

To improve the performance based on the VQC method in OpenAI Gym tasks, we propose a new architecture, SVQC, {which is} 
composed of three parts: input, parameter, and output layers shown in Fig.~\ref{Fig_ReuseCircuit}. For input layer, the environment provides an 
{input} state {$x_i$} to the input layer. The state {$x_i$} is encoded by the angle of a {single-qubit} rotational gate $\mathrm{U}(x_i)$. 
{The number of the data encoding rotational gates is usually determined by the state dimension.} 


In the parameters layer, trainable parameters $\theta_i$ control the {single-qubit} rotational gates $\mathrm{U(\theta_i)}$ and are updated through gradient descent. Here, we only use the single-qubit gates without entangling gates to overcome the problem of the barren plateau caused by entangling gates in the optimization process~\cite{PRXQuantum.2.040316}.

The output layer is decomposed by three components: {measurements, connection with a classical NN, and output reuse strategies}.
First, the expectation value of the measurements is obtained from Eq.~(\ref{general_circuitoutput}) {denoted by
\begin{equation}
\mathrm{y}_{i q}=\mathrm{f}(x_{i}^q, \theta_{i}^q) =\left\langle 0\left|\operatorname{U}^{\dagger}(x_{i}^q, \theta_i^q) \operatorname{Z}_q \operatorname{U}(x_{i}^q, \theta_i^q)\right| 0\right\rangle\label{NotrainOutput},
\end{equation}
where $q=1,\dots,n$ is the indexes for different single qubits in the quantum circuit, $\operatorname{Z}_q$ is the Pauli Z matrix of qubit $q$, and $\operatorname{U}(x_{i}^q,\theta_{i}^q) \in \mathbb{C}^{2 \times 2}$ is a single-qubit unitary operator.} Since the SVQC outputs merely come from the expectation values of eigenvalues on the unitary family $\{\operatorname{U}^{\dagger}(x_{i}^q, \theta_i^q) \operatorname{Z}_q \operatorname{U}(x_{i}^q, \theta_i^q)\}$, the following technical strategies can enrich the expressive power of its outputs. Second, we link the {single-qubit-based} quantum circuit with a fully-connected NN layer to increase the expressive power of the quantum circuit so that it is more likely to achieve the optimal result in the practical optimization process~\cite{Parametrized_quantum_policies_Jerbi, PhysRevLett.127.090506}. Given that there exist optimal circuit outputs {$\mathrm{y}_{i p}^{\prime *}(x_{i}^q, \theta_{i}^q) \in \mathbb{R}$ for variety tasks on OpenAI Gym, our target is to minimize $|\mathrm{y}_{i p}^{\prime}-\mathrm{y}_{i p}^{\prime *}|$, where $\mathrm{y}_{i p}^{\prime} \in \mathbb{R}$ is the scaled quantum circuit output for $p=1,\dots,k$:} 
\begin{equation}
{\small
\left(\begin{array}{c}
\mathrm{y}_{i 1}^{\prime} \\
\vdots \\
\mathrm{y}_{i p}^{\prime}  \\
\vdots \\
\mathrm{y}_{i k}^{\prime} 
\end{array}\right)=\left(\begin{array}{c}
{b}_{i 1}  \\
\vdots \\
{b}_{i p}  \\
\vdots \\
{b}_{i k} 
\end{array}\right)+\left(\begin{array}{ccccc}
W_{11}^i & \cdots & W_{1 q}^i& \cdots & W_{1 n}^i \\
\vdots & \ddots& \vdots & \ddots & \vdots \\
W_{p1}^i& \cdots& W_{pq}^i & \cdots & W_{pn}^i \\
\vdots & \ddots& \vdots & \ddots & \vdots \\
W_{k 1}^i & \cdots & W_{kq}^i & \cdots &W_{k n}^i
\end{array}\right)\left(\begin{array}{c}
\mathrm{y}_{i 1} \\
\vdots \\
\mathrm{y}_{i q} \\
\vdots \\
\mathrm{y}_{i n}
\end{array}\right)\label{trainableoutput}}, 
\end{equation}
{where $W_{p q} \in \mathbb{R}$ is the trainable parameters (weights) in the NN, $ b_{i p} \in \mathbb{R}$ are the biases,
$k$  is the number of actions and $n$ is the number of qubits. Comparing the domain of $\mathrm{y}_{i q}$ with that of $\mathrm{y}_{i p}^{\prime}$} in Eqs.~(\ref{NotrainOutput}) and~(\ref{trainableoutput}), the latter can increase the expressive power of the circuit output according to the studies in Ref.~\cite{Parametrized_quantum_policies_Jerbi, PhysRevLett.127.090506}. 

\begin{table*}[htbp]
  \caption{\label{tab_deataillearning} {Details of model settings for different RL environments.
  The quantum circuits shown in Fig.~\ref{Fig_ReuseCircuit} are employed} to complete the different {RL} tasks. The initial state is set to be in the ground state while the input layer is composed of a Hadamard gate followed by the gate sequence of {(Ry($x^{q}_{i}$)-Rz($x^{q}_{i}$), where $q=1, \ldots,n$ are the qubit indexes. The parameterized single-qubit gate is Ry($\theta^{q}_{i}$). The number of qubits for the  CartPole, Acrobot, and LunarLander environments are $n$ = 4, 6, and 8, respectively. Then the repeated measurements of $\sigma_{z}^{q}$ are performed in the output layer. Finally, the measurement outputs could be reused for certain times
and then connected with a fully-connected classical NN layer. The numbers appearing in} the parentheses of FCN and CNN are the numbers of nodes in the fully-connected classical NN layers.}
 
  \begin{ruledtabular}
    \begin{tabular}{c c c c c c  c}
     Environment  & Learning algorithm &  Architecture ({qubit number $n$}) &  {Times of reuses} ($\ell$) & Number of actions ($k$) \\
    \hline
      CartPole-v1 &Algorithm~\ref{alg:1} & {$|0\rangle$-H-Ry($x^{q}_{i}$)-Rz($x^{q}_{i}$)-Ry($\theta^{q}_{i}$), $n=4$} & 16 & 2\\
      CartPole-v1  &Classical PPO& FCN (16, 32, 64, 32, 2) & None 
      \\
      CartPole-v1  &Classical PPO& CNN (5, 2, 4, 2) & None 
      \\      
      Acrobot-v1  & Algorithm~\ref{alg:1}  & {$|0\rangle$-H-Ry($x^{q}_{i}$)-Rz($x^{q}_{i}$)-Ry($\theta^{q}_{i}$), $n=6$} & 8 &3  \\
      Acrobot-v1 &Classical PPO& FCN (16, 32, 64, 32, 3) & None 
      \\
      Acrobot-v1 &Classical PPO& CNN (5, 2, 4, 2, 3) & None 
      \\
      LunarLander-v2  &Algorithm~\ref{alg:1}, ~\ref{alg:2}& \begin{tabular}[c]{@{}c@{}}{$|0\rangle$-H-Ry($x^{q}_{i}$)-Rz($x^{q}_{i}$)-Ry($\theta^{q}_{i}$)}, $n=8$\\  {repeated} 3 times\end{tabular}
       & 8 &4 \\
      LunarLander-v2 &Classical PPO& FCN (16, 32, 64, 32, 4) & None 
      \end{tabular}%
  \end{ruledtabular}


\end{table*}

\begin{figure}[http]
\begin{algorithm}[H]
  \caption{Hybrid Quantum PPO (QPPO) algorithm}
  \label{alg:1}
     \hspace*{\algorithmicindent} \textbf{Input}: State: $(x_1, s_2,\dots,x_{i})$\\
\hspace*{\algorithmicindent} \textbf{Output}: Action: $a_{i}$ \\
   \begin{algorithmic}
\For{episode}
    \State{Transfer from classical state $x_{i}$ to quantum stat $|\psi\left(x_{i}\right)\rangle$}
    \State{Actor collect data $\operatorname{D} (s, a, r, \mathrm{P_a})$ through VQC} 
    \State{actor’s policy $\pi_{\theta_a}\left(s\right)$ from environment.}
    \State{Actor VQC output $\mathrm{P_a}$ and action $a$.}
    \State{Memorize the $\left(s, a, r, \mathrm{P_a}\right)$ into experience buffer}
    \State{Critic VQC output the value$ \mathrm{V}\left(s\right)$ through state $s$}
    \State{Compute discount reward $R_t=\sum_{t=1}^{T}\gamma^{t+T-1} \mathrm{r}$}
    \State{Compute advantage estimate $a_i=R_t-V\left(x_i\right)$}
    \If{step \% update-time == 0 or done}
        \For{epoch}
            \State{Update critic parameters $\theta_c$ through minimum} 
            \State{value loss $L_v=\frac{1}{\left|D\right|}\sum_{t=1}^{T}\left(V\left(x_i\right)-R_t\right)^2$} 
            \State{by gradient descent with Adam.}
            \State{Update actor parameters $\theta_a$ through maximum} 
            \State{actor loss.}  \State{$\mathrm{L}_a=\frac{1}{\left|D\right|}log\sum_{t=1}^{T}{\min(\frac{\pi_\theta(a_i|x_i)}{\pi_{\theta_{old}}(a_i|x_i)}}$,}
            \State{$clip\left(1+\epsilon,1-\epsilon\right)) \mathrm{A}_{t}$}
            \State{use gradient ascent with Adam.}
            \State{Update VQC actor and classical NN parameters} \State{$\theta_{a}\longleftarrow\ \theta_{a}+\nabla_{\theta_{a}}\mathrm{L}_{a}$.}
            \State{Clean Experience}
        \EndFor
    \EndIf
\EndFor
   \end{algorithmic}
\end{algorithm}
\end{figure}
\begin{figure}[http]
\begin{algorithm}[H]
  \caption{Hybrid policy on LunarLander-v2 environment.}
  \label{alg:2}
   \begin{algorithmic}[1]
   \If{episode reward $>=$ 200}
    \State{Stop update parameters}
    \State{count $+=$ 1}
    \If{count $>$ Max conut}
        \State{Max conut $=$ count}
        \State{Save actor and critic policy}
        \EndIf
    \If{average episode reward $<$ 200}
        \State{Update actor and critic parameters}
        \State{count $=$ 0}
        \EndIf
    \EndIf
   \end{algorithmic}
\end{algorithm}
\end{figure}

Finally, we duplicate the expectation value of the measurement $\ell$ times and then all outputs are fed into the classical fully-connected NN layer shown in Fig.~\ref{Fig_ReuseCircuit}. 
{The final scaled output with the duplicated qubit outputs reads
\begin{equation}
    \mathrm{y}_{i p}^{\prime}={b}_{i p}+{W}_{p 1}^{i\#}\mathrm{y}_{i 1}^{\prime}+\dots+{W}_{p q}^{i\#}\mathrm{y}_{i q}^{\prime}+\dots+{W}_{p n}^{i\#}\mathrm{y}_{i n}^{\prime},
    \label{reuse_equation}
\end{equation}
where $W_{k j}^{i\#}=W^{i(1)}_{k j}+W^{i(2)}_{k j}+\cdots+W^{i(\ell)}_{k j}$ is the expected weights with duplicated outputs}, $n \in \mathbb{N}$ {denotes} the $n\text{th}$ qubit, and $\ell \in \mathbb{N}$ is the number of {the output reuse.
It will be shown later that the method of reusing qubit outputs improves the sample efficiency.}

\begin{figure}[http]
  \centering
  \includegraphics[width=1\linewidth]{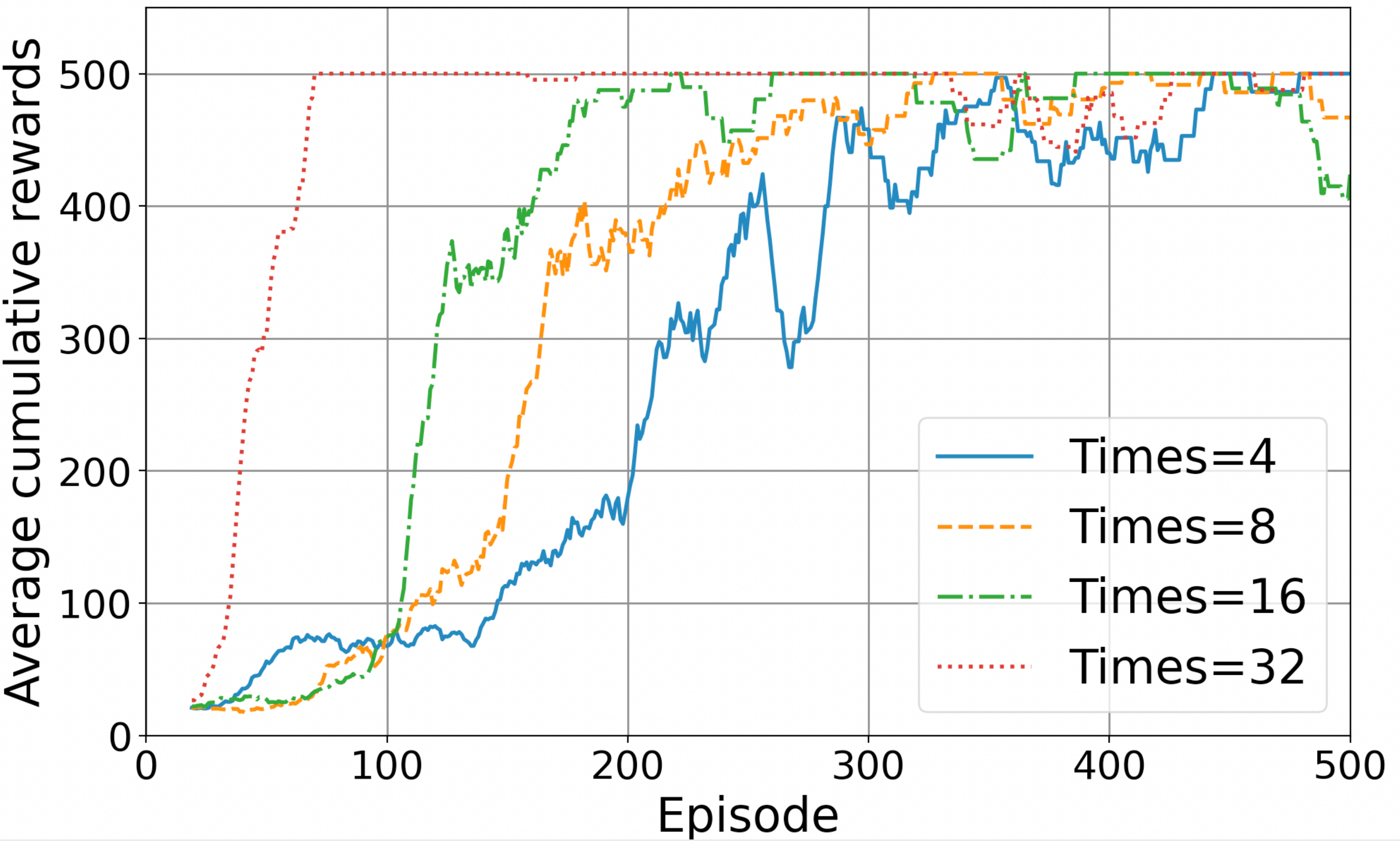}
  \caption{{Performances of the quantum RL agents for different numbers of output reuse of   4, 8, 16 and 32 times for the CartPole environment. The x-axis is the number of episodes, and the y-axis is the average cumulative reward, averaged over the last 20 episodes.} 
  }
  \label{fig_reuse}
\end{figure}

\begin{table*}[htbp]
\centering
  \caption{Hyperparameters of different models}
  \begin{ruledtabular}
    \begin{tabular}{p{8em} c c c c c c c }
      Environment  & Learning algorithm & Architecture & Actor Learning rate  & Critic Learning rate & Discount factor & epoch & Clip $\epsilon$  \\
      \hline
      CartPole-v1 & VQC-PPO Alg.~\ref{alg:1} & Fig.~\ref{Fig_ReuseCircuit} & 0.001 & 0.01  & 0.99  & 4     & 0.1 \\
       CartPole-v1 & VQC-PPO Alg.~\ref{alg:1} & Fig.~\ref{Fig_IBM_architecture_detail}(a) &0.004 & 0.04  & 0.99  & 4     & 0.1 \\
      CartPole-v1 & Classical PPO & Neural network & 0.0003 & 0.001 & 0.98  & 4     & 0.1 \\
      Acrobot-v1 & VQC PPO Alg.~\ref{alg:1} & Fig.~\ref{Fig_ReuseCircuit} & 0.004 & 0.04  & 0.98  & 4     & 0.1 \\
      Acrobat-v1 & Classical PPO  &Neural network & 0.0003 & 0.001 & 0.98  & 4     & 0.1 \\
      LunarLander-v2 & VQC PPO Alg.~\ref{alg:1}~\ref{alg:2} & Fig.~\ref{Fig_ReuseCircuit}&0.002 & 0.02  & 0.98  & 4     & 0.1 \\
      
      \end{tabular}%
      \label{tab_hyperparameters}
  \end{ruledtabular}
\end{table*}

\section{\label{sec:4s}Numerical results and discussion}

To improve sample inefficiency in OpenAI Gym tasks,
we compare the learning curves of SVQC, 
MVQC, and classical NNs on different tasks.
Moreover, we use the IBM quantum devices to test the CartPole, Acrobot, and LunarLander tasks for comparing the real devices with {an} ideal simulator. 
In the following, the detailed discussion of simulator results are shown in Section~\ref{Result_on_simulator} and the results of IBM quantum devices are shown in Section~\ref{Result_on_NISQ_device}.
The {details of model settings including circuit architectures and learning algorithms of SVQC, classical fully-connected neural network (FCN), and convolution neural network (CNN) for simulations of different RL environments} on simulators are {described} in Table~\ref{tab_deataillearning}.
The detailed hyperparameters of {the simulation} models are shown in Table~\ref{tab_hyperparameters}.



\subsection{\label{Result_on_simulator} Results on simulator}

\begin{figure*}[htbp]
  \centering
  \includegraphics[width=1\linewidth]{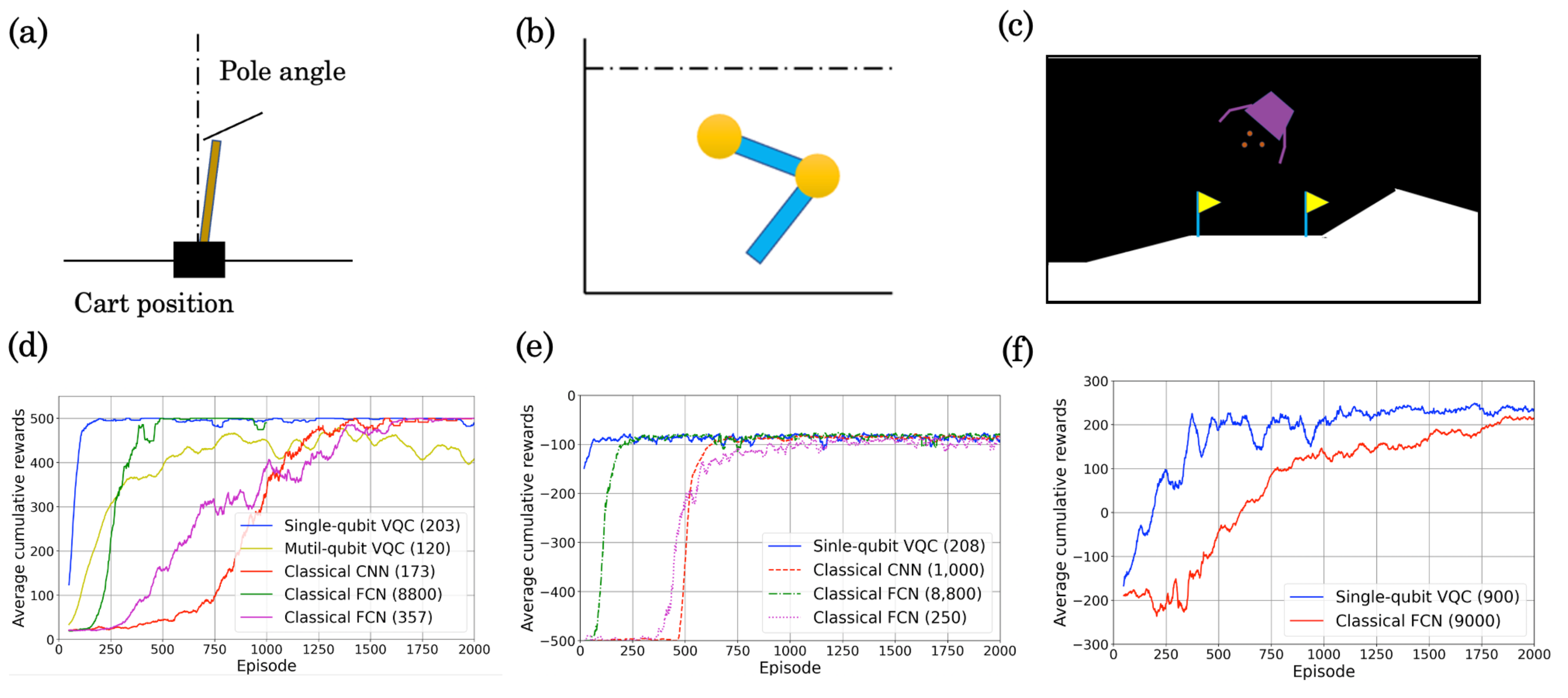}
  \caption{{Illustrations and Performances} of the single-qubit VQC PPO using single-qubit systems with output reuse strategy, classical fully-connected neural network (FCN) and convolutional neural network (CNN) in {the RL} optimization process for (a)(d) CartPole, (b)(e) Acrobot, and (c)(f) LunarLander problems. The {numbers in the parentheses of different models are the total trainable parameters used in these models, respectively.}
  The x-axis is the number of {episodes and the y-axis represents the average cumulative reward, averaged over the last 20 episodes, at that episode.} 
  The CartPole and Acrobot experimental results are averaged over five runs and the LunarLander result is the best one in 10 runs. 
  The sample efficiency follows {approximately the relationship of} SVQC $>$ classical FCN $\approx$ CNN $>$ Multi-qubit VQC.}
  \label{fig_openAIresult}
\end{figure*}

\begin{table*}[htbp]
  \caption{{Relaxation time $T_1$, dephasing time $T_2$, single-qubit $\sqrt{\mathrm{X}}$, X, and  identity (ID) gate errors and readout error data of different quantum machines downloaded from IBM Quantum service at the time when the experiments were performed.}}
  \begin{ruledtabular}
    \begin{tabular}{cccp{4em}<{\centering}cccp{5em}<{\centering}}
      Machine & $\mathrm{T}_{1}$ (us) & $\mathrm{T}_{2}$ (us) & Readout assignment error & Readout length (ns) & ID error &$\sqrt{\mathrm{X}}(\mathrm{S_x})$ error & Single-qubit Pauli-X error \\
      \hline
      ibmq\_lagos & 75.65 & 39.3  & 1.16E-02 & 704   & 3.11E-04 & 3.11E-04 & 3.11E-04 \\
      Ibmq\_belem & 86.5  & 100.93 & 2.46E-02 & 5351.111 & 2.33E-04 & 2.33E-04 & 2.33E-04 \\
      Ibmq\_lima & 87.9  & 87.83 & 2.49E-02 & 5351.111 & 5.31E-04 & 5.31E-04 & 5.31E-04 \\
      ibmq\_jakarta & 91.96 & 41.46 & 3.41E-02 & 5351.111 & 3.84E-04 & 3.84E-04 & 3.84E-04 \\
      ibmq\_toronto & 92.3  & 55.67 & 4.57E-02 & 5201.778 & 2.34E-04 & 2.34E-04 & 2.34E-04 \\
      \end{tabular}%
  \end{ruledtabular}
  \label{ap_quantum_devices_property}
\end{table*}

\begin{figure*}[htbp]
  \centering
  \includegraphics[width=1\linewidth]{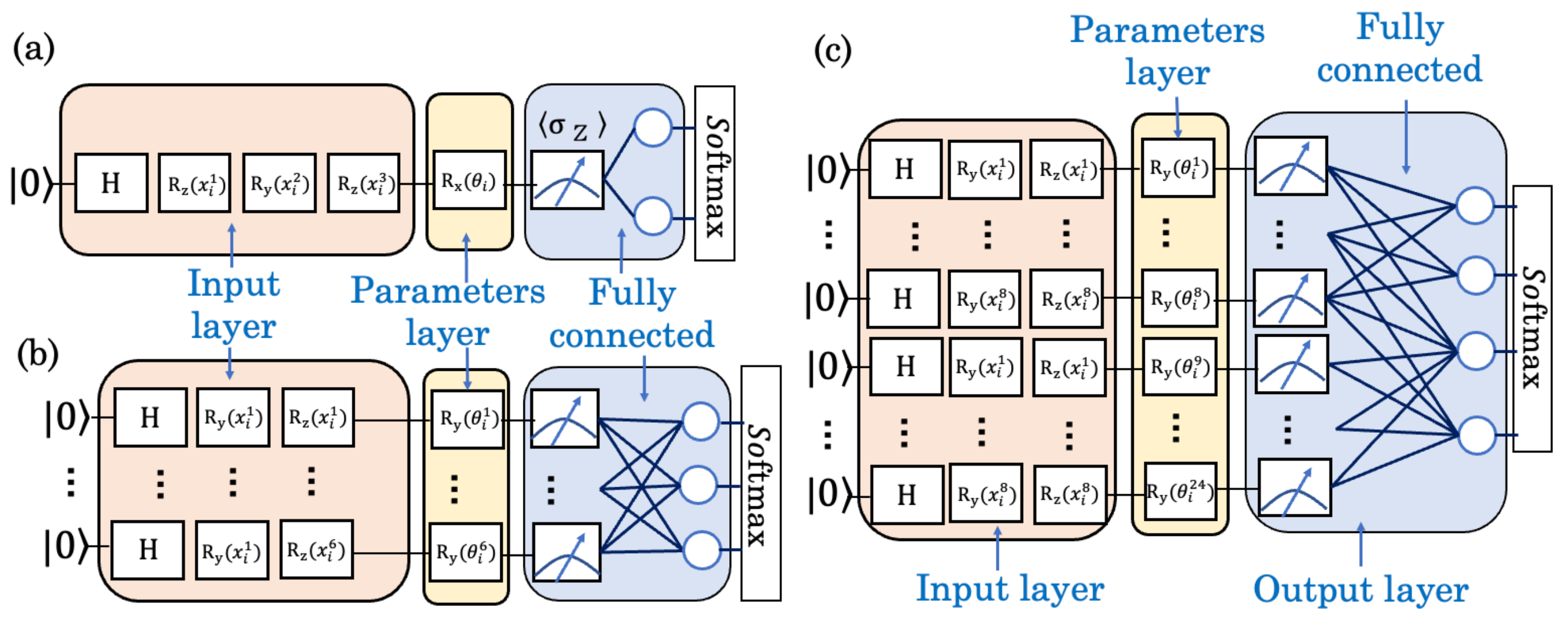}
  \caption{Architectures of the SVQC models for the (a) CartPole-v0, (b) Acrobat-v1 and (c) LunarLander-v2 task for the tests on real quantum devices and a simulator. H {stands for} the Hadamard gate. {$x^{j}_{i}$ is the $j\mathrm{th}$ feature state on the} $i\mathrm{th}$ episode.  {$\theta^q_{i}$ is the trained parameters on the $q\mathrm{th}$ qubit in the quantum circuit on the $i\mathrm{th}$ episode.}}
  \label{Fig_IBM_architecture_detail}
\end{figure*}

\begin{figure*}[htbp]
  \centering
  \includegraphics[width=1\linewidth]{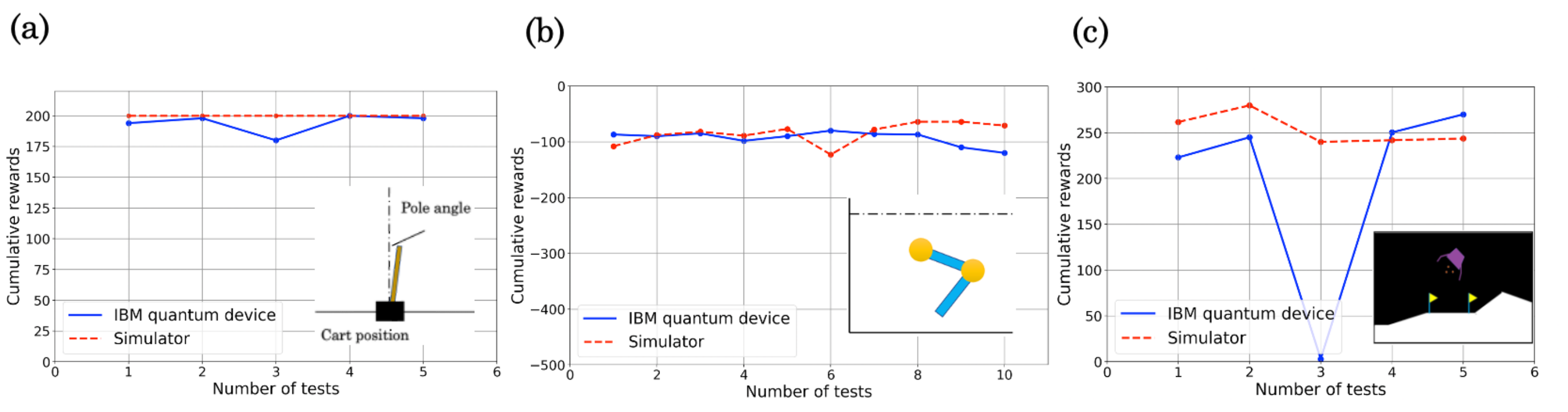}
  \caption{Performances of {the tests of the SVQC RL agents for the} (a) CartPole-v0, (b) Acrobat-v1 and (c) LunarLander-v2 tasks on IBM quantum devices and a simulator. The x-axis is the {index} number of the tests. The y-axis is the cumulative rewards in one episode.
  The mean rewards of CartPole, Acrobot and LunarLander on {the} quantum devices are 194.0, -90.04, and 198.3, and the standard deviations are 8.12, 5.03, and 19.3, {respectively}. The average rewards of CartPole, Acrobot and LunarLander on {the} ideal simulator are 200.0, -84.40, and 253.4, and the standard deviations are 0, 18.8, and 17.1,     {respectively}.
  }
  \label{Fig_IBM_quantumResult}
\end{figure*}

{The training processes of SVQCs in {the CartPole and Acrobot {tasks} are attained through Algorithm~\ref{alg:1},} while both Algorithm~\ref{alg:1} and Algorithm~\ref{alg:2} are used in {the} LunarLander-v2 task. 
{The performances for different numbers of the reused outputs for the CartPole environment is shown in Fig.~\ref{fig_reuse}, indicating that the method of reusing qubit outputs improves} the sample efficiency. 
The learning curves of the various models on the tasks of CartPole, Acrobot, and LunarLander are shown in Fig.~\ref{fig_openAIresult} (d)-(f), respectively.}


For the CartPole task, we compare the sample efficiency of single-qubit VQC (SVQC), multi-qubit VQC (MVQC), classical FCN, and CNN {in Fig.~\ref{fig_openAIresult} (d)}. 
{Our work uses the SVQC architecture shown in Fig.~\ref{Fig_ReuseCircuit} and described in Table~\ref{tab_deataillearning}. The MVQC model {for RL} is discussed in some detail in
$\href{https://www.tensorflow.org/quantum/tutorials/quantum_reinforcement_learning}{\text{tensorflow-quantum}\text{ }\text{tutorial}}$ and the circuit architecture of MVQC used here is from~\cite{Parametrized_quantum_policies_Jerbi}. Descriptions about the architectures of classical FCN and CNN used here can be found in Table~\ref{tab_deataillearning}}.
{From Fig.~\ref{fig_openAIresult} (d), we find that} SVQC (thick solid line) achieves maximum rewards {in} about 150 episodes. In comparison, MVQC (dashed line) converges around 500 episodes without achieving the maximum rewards {while} FCN (8,800) (dash-dot line), FCN (357) (dotted line), and CNN (173) reach the maximum rewards in about 400, 1,600, and 1650 episodes, respectively. {Note that the numbers in the parentheses of different models are the total trainable parameters used in these models, respectively.}
{These results provide concrete numerical evidence that} SVQC could improve sample efficiency {by} {about} three times compared to {classical FCN (8,800)} under the same optimization process.

For the Acrobot task, we compare the sample efficiency of SVQC, classical fully-connected neural network (FCN), and convolution neural network (CNN) {in Fig.~\ref{fig_openAIresult} (e)}. From the figure, we find SVQC (solid line) achieves the average rewards of -90 in around 90 episodes while FCN (8,800) (dashed line), FCN (250) (dash-dot), and CNN (1,000) (dash-dot) reach the average reward of -90 {in} about 250, 600, and 1000 episodes. These results also indicate that SVQC could improve the sample efficiency by about three times compared to { classical FCN (8,800)} under the same optimization process.

{For the Luarlander task}, we compare the sample efficiency of SVQC and FCN in Fig.~\ref{fig_openAIresult} (f). The figure shows that SVQC (solid line) achieves the average rewards of about 220 in 1,000 episodes. In comparison, FCN (dashed line) reaches the average rewards of 200, slightly lower than SVQC, in 1,750 episodes. These results conclude that the SVQC improves the sample efficiency by about two times compared to {classical FCN (9,000)} under the same optimization process. {We would like to emphasize that this is the first time that the {more complex} control task of LunarLander can be achieved in {the} quantum RL field.}

{In conclusion, our proposed SVQC achieves higher rewards than existing VQC-based models~\cite{Lockwood_Si_2020, Parametrized_quantum_policies_Jerbi, skolik2021quantum, 9620885} and improves the {sample efficiency (speed of convergence)} compared to the classical fully-connected neural networks in {the} CartPole and Acrobot tasks.} {Moreover, 
our SVQC uses much (at least one order of magnitude) fewer trainable parameters with even better learning performance than the best performing FCNs shown in Fig.~\ref{fig_openAIresult} (d)-(f)}. 


\subsection{\label{Result_on_NISQ_device}Implementation on IBM quantum devices}
In this section, we feed the trained parameters {of our SVQCs} into the IBM quantum devices, and compare their performance with that of the ideal simulators. The chosen benchmarking environments are CartPole-v0, Acrobot-v1, and LunarLander-v2 in OpenAI Gym. {The {real quantum device experiments for the} CartPole and Acrobot tasks are executed on ``ibm\_lagos" and ``ibm\_belem",  while the LunarLander task is on ``ibm\_lima", ``ibm\_jakarta" and ``ibm\_toronto".}

The cumulative rewards on {the} CartPole-v0, Acrobot-v1, and LunarLander-v2 tasks {are} shown in Fig.~\ref{Fig_IBM_quantumResult} (a)-(c) {and the numbers of measurements for these tasks are} 1024, 1024, and 8192, respectively. Details of the used hyperparameters can be found in Table.~\ref{ap_quantum_devices_property}, and we will elaborate on the circuit architectures and their performances below.

We employ the quantum circuit in Fig.~\ref{Fig_IBM_architecture_detail} (a) to conduct the task of CartPole-v0, where only a single qubit initialized in the ground state, $|0\rangle$, is required. The input layer consists of a Hadamard gate and the sequence of (Rz, Ry, Rz) gates; while the parameter layer contains only an Rx gate. 
{We import four classical trained parameters to scale the circuit output and use a {trainable} parameter for the angle of Rx.}
{We upload the
trained model parameters of the SVQC model in Fig.~\ref{Fig_IBM_architecture_detail} (a) to IBM Quantum devices.} 
According to Fig.~\ref{Fig_IBM_quantumResult}(a), we find that the average reward of the real device over five tests is 190. This reward is comparable with the average reward, 200, obtained {on} the ideal simulator. 

{The six-qubit quantum circuit shown in Fig.~\ref{Fig_IBM_architecture_detail} (b) is used to complete the Acrobot-v1 task. Similarly, the initial state is set to be in the ground state while the input {and parameter layers are} composed of a Hadamard gate followed by the gate sequence of (Ry, Rz, Ry). Then the repeated measurements of $\sigma_{z}$ are performed in the output layer. This circuit consists of six trainable rotational angles and 21 parameters for output rescaling. The average reward obtained from the quantum machine is -90 while the value from the ideal simulator is -84. Again, the {rewards} of the real device and the idea simulator {are} comparable. }


{The first quantum RL agent based on SVQC consisting of a 24-qubit quantum circuit with the architecture demonstrated in Fig.~\ref{Fig_IBM_architecture_detail} (c) is used to complete the LunarLander-v2 task on the real quantum device. 
The circuit inherits the architecture of SVQC used in the Acrobot-v1 task and is further expanded to a larger scale by introducing 24 trainable angle parameters and 100 output scaling parameters. By comparing the average reward of the real device with that of the ideal simulator in Fig.~\ref{Fig_IBM_quantumResult} (c), we find the average reward of the real device is 200, 
{slightly worse than the average reward of 250 of}
the ideal simulator. 
We expect that conducting more test runs will further increase the average reward of the real device, but we did not continue the experiments on IBM devices due to the {cost of the expensive quantum} computational resources. 
}



To the best of our knowledge, it is the first time the complex RL tasks can be accomplished on IBM quantum devices. The results demonstrate that even though the training is on the quantum simulator without noise, the trained SVQC models have similar performances on current NISQ devices compared to the ideal simulator.

\section{Conclusion and future work}\label{sec:5}

The SVQC performs better than previous studies~\cite{Lockwood_Si_2020, Parametrized_quantum_policies_Jerbi, skolik2021quantum, 9620885} which use several CNOT or CZ gates on the CartPole and Acrobot {environments}. This leads to an open question of~\emph{``What are the roles of entangling gates in MVQC for the RL tasks?}'' On the other hand, we find SVQC with the output reuse can solve the RL tasks more efficiently than the classical NNs. This brings the question of ``\emph{What is the quantum-inspired algorithm, which can solve the RL problems efficiently?}'' Since SVQC can be implemented on the current NISQ quantum devices to handle classical control and box2d tasks in openAI Gym. Therefore, ``\emph{What is the limitation on the current NISQ devices in RL tasks?}'' is the remaining question for the future work.


\section{Acknowledgments}
J.Y.H and H.S.G. thank IBM Quantum Hub at NTU for providing computational resources and accesses for conducting the real quantum machine experiments.
H.S.G. acknowledges support from the the Ministry of Science and Technology
of Taiwan under Grants No.~MOST 109-2112-M-002-023-MY3, 
No.~MOST 109-2627-M-002-003, No.~MOST 110-2627-M-002-002,
No.~MOST 107-2627-E-002-001-MY3, No.~MOST 111-2119-M-002-006-MY3
and No.~MOST 110-2622-8-002-014
from the US Air Force Office of Scientific Research under
award number FA2386-20-1-4033,
and from the
National Taiwan University under Grant
No.~NTU-CC-111L894604.\\

Code availability: 
The Codes that support the findings of this study and all trained parameters in different tasks are available {at}
$\href{https://github.com/Yueh-H/single-qubit-RL}
{\text{https://github.com/Yueh-H/single-qubit-RL}}$.

\bibliographystyle{apsrev4-2}
\bibliography{sorsamp}

\appendix

\section{Supplementary materials of variational quantum circuits and the environments}\label{Appendix:VQC_Environments}
In this Appendix, we discuss the variational quantum circuit (VQC) in Sec.~\ref{appendix_VQC}, the architecture of VQC-based quantum reinforcement learning in Sec.~\ref{appendix_VQC_RL}, and the detailed constraints on different environments on OpenAI Gym in Sec.~\ref{appedix_introduction_to_openAI_gym}.



\subsection{\label{appendix_VQC}Discussion of variational quantum circuits}

When it comes to quantum computation, the goal is to find the quantum advantage. The potential advantage in VQC is to use the vast Hilbert space in the quantum information process~\cite{PhysRevLett.122.040504, Biamonte2017}. There are several steps of VQC to explore the potential of {the} advantage. Let the data set $\mathcal{D}={\{x_1,\dots,x_M\}}$ 
{has $M$ feature data and each is an  $N$-dimensional real feature vector}.

\begin{enumerate}
\item State preprocess layer:

There are three main encoding strategies of the quantum circuit{: basis, amplitude~\cite{18, 19}, and Hamiltonian encoding schemes. The basis encoding scheme needs the runtime of} $\mathcal{O}(MN)$ for state preparation without QRAM~\cite{PhysRevLett.100.160501}, while the amplitude and Hamiltonian encoding schemes can reduce the time to $\mathcal{O}(\log(MN))$ with QRAM.
Moreover, the Hamiltonian encoding tries to build the kernel space, which is hard to be built using classical computers~\cite{2019_nature}.

\item Parameter layer:

There are different technologies to design the different architectures of the circuit by machine learning~\cite{du2020quantum} or reinforcement learning~\cite{kuo2021quantum, ye2021quantum}.

\item Measurement:

The general quantum circuit output is $\left\langle\sigma_{z}\right\rangle=\operatorname{tr}\left(\operatorname{\rho}(x_{i},\theta_{i}) \operatorname{\sigma_{z}}\right)$, where $\rho(x_{i}, \theta_{i}) \in \mathbb{C}^{2^n \times 2^n} $ is the density matrix depending on parameters and input data, and $\sigma_{z} \in \mathbb{C}^{2^n \times 2^n}$ is the projective matrix. A challenge about {the} measurement is that lots of shots would eliminate the runtime advantage~\cite{2018_classicalperspective, Aaronson2015QuantumML}. There are strategies to improve the efficiency in measuring the quantum state~\cite{du2021efficient, PhysRevResearch.3.043095}.

\item Optimization:

The challenge about circuit optimization lies in barren plateau~\cite{McClean2018}. The gradient of parameters would vanish exponentially in the optimization process. Using tree structure~\cite{zhang2020trainability}, tuning the parameters with an iterative optimization structure, and using adaptively selected Hamiltonian~\cite{liao2021quantum_optimization} can mitigate the barren plateau in the process.
\end{enumerate}

\subsection{Discussion of VQC-based quantum reinforcement learning }\label{appendix_VQC_RL}

There are many technical skills in VQC-based quantum reinforcement learning. {References} \cite{Parametrized_quantum_policies_Jerbi, skolik2021quantum, lan2021variational_soft_actor_critic} {provide} various methods to solve the OpenAI Gym tasks. The methods can be divided by the circuit architecture that consists of the input, parametric, and output layers.

In the input layer, the additional trainable parameters are encoded by the rotational {angles of the gates that improve} the performance on the Cartpole and Acrobot tasks~\cite{Parametrized_quantum_policies_Jerbi, skolik2021quantum}. For the input and parametric layers, the repeated application of re-uploading enhances the performance on the classical control tasks~\cite{Parametrized_quantum_policies_Jerbi, skolik2021quantum, lan2021variational_soft_actor_critic}. In the output layer, introducing the extra trainable parameters to rescale the measurement outcomes~\cite{Parametrized_quantum_policies_Jerbi, skolik2021quantum} or adding the classical neuron network connection improves the cumulative rewards~\cite{lan2021variational_soft_actor_critic} on the Cartpole and Pendulum tasks.

\begin{table}[htbp]
  \caption{\label{TabCarpole_introduction}{Constrains on the observations of the} Cartpole environment. The {termination} condition is that the pole exceeds 12 degrees or the Cart position exceeds 2.4 or -2.4.}
  \begin{ruledtabular}
    \begin{tabular}{ccc}
      Observation	& Min	& Max \\
      \hline
      Cart Position $x$ & 	-4.8 & 	4.8 \\ 
      Cart Velocity $v$	& $-\mathrm{Inf}(-\infty)$	& $\mathrm{Inf}\ (+\infty)$\\ 
      Pole Angle $\theta$	& -0.418 rad	& 0.418 rad \\ 
      Pole Angular Velocity $\theta$	& $-\mathrm{Inf}\ (-\infty)$ &$	\mathrm{Inf}\ (+\infty)$
    \end{tabular}
  \end{ruledtabular}
\end{table}

\begin{table}[htbp]
  \caption{\label{TabAcrobot_introduction}{Constrains on the observations of the} Acrobot environment. The episode terminates when the {end of the lower link exceeds the given height} or the agent does not achieve the condition within 500 {time} steps.}
  \begin{ruledtabular}
    \begin{tabular}{ccc}
      Observation & Min   & Max \\
      \hline
      upper pole cos & -1    & 1 \\
      upper pole sin  & -1    & 1 \\
      down pole cos & -1    & 1 \\
      down pole sin & -1    & 1 \\
      Upper angular velocity &      -4$\pi$ & 4$\pi$ \\
      down angular velocity &      -9$\pi$ &  9$\pi$\\
      \end{tabular}
  \end{ruledtabular}
\end{table}

\subsection{\label{appedix_introduction_to_openAI_gym}Introduction to the OpenAI Gym environment}

The followings are the constraints {on the} Cartpole, Acrobot, and LunarLander environments. The number of {states} of Cartpole , Acrobot, and LunarLander are four, six, and eight{, respectively, and the numbers of actions are} two, three, and four{, respectively.} The {constrains on different observations (states)} of Cartpole and Acrobot are subsequently shown in Table.~\ref{TabCarpole_introduction} and Table.~\ref{TabAcrobot_introduction}.

The detailed information of LunarLander is as follows. According to the description of the environment in OpenAI Gym, the reward for moving from the top of the screen to the landing pad and zero speed falls between 100 and 140 points. The episode finishes if the lander crashes or comes to rest, receiving an additional reward of $-100$ or $+100$ points. Each leg ground contact is $+10$.
The reward is −0.03  for firing the side engine, and −0.3 for firing the main engine each frame.

\end{document}